\documentstyle[prl,floats,twocolumn,epsf,psfrag,aps]{revtex}

\tighten
\begin{document}
\draft
\twocolumn[\hsize\textwidth\columnwidth\hsize\csname @twocolumnfalse\endcsname

\title{Multipartite entanglement for continuous variables:
A quantum teleportation network}
\author{P.\ van Loock and Samuel L.\ Braunstein}
\address{Quantum Optics and Information Group,\\
School of Informatics, University of Wales, Bangor LL57 1UT, UK}
\maketitle

\begin{abstract}
We show that {\it one\/} single-mode squeezed state 
distributed among $N$ parties using linear optics suffices to produce 
a truly $N$-partite entangled state for any nonzero squeezing and
arbitrarily many parties.
From this $N$-partite entangled state, via quadrature measurements
of $N-2$ modes, bipartite entanglement between any two of the $N$ 
parties can be `distilled', which enables quantum teleportation with 
an experimentally determinable fidelity better than could be 
achieved in any classical scheme.
\end{abstract}
\vspace{3ex}
]

Entanglement is seen as an essential ingredient in quantum communication 
and computation. For example, it enables quantum teleportation which
was originally proposed for systems of discrete variables \cite{Benn1}. 
Later, quantum teleportation was also proposed for continuous 
variables \cite{Vaid,Sam98a}.
The simplest teleportation schemes rely on bipartite entanglement, the 
entanglement of a pair of systems shared by two parties. For pure states, 
this kind of entanglement is well-understood and can be quantified 
\cite{Popes}. Multipartite entanglement, the entanglement shared by more 
than two parties, is much more difficult to quantify \cite{Thap}.
Yet in the laboratory, the creation of tripartite discrete-variable 
entanglement, yielding so-called GHZ states \cite{GHZ},
has been reported for single-photon polarization states \cite{Bouw} and 
using nuclear magnetic resonance \cite{Lafl}.

Continuous-variable quantum teleportation of arbitrary
coherent states has been realized experimentally 
with bipartite entanglement built from two single-mode squeezed vacuum 
states combined at a beamsplitter \cite{Furu}.
In the absence of entanglement the best mean fidelity of
the reconstructed coherent states is $F=\case{1}{2}$ \cite{Fuchs}. 
Experimentally, 
$F=0.58\pm 0.02$ was achieved. Though this limits our attention to
the teleportation of a rather modest set of non-orthogonal states, 
{\it the fidelity gives a clear experimental signal for the
presence of entanglement}.

Now it is known that even one single-mode squeezed state incident on a 
beamsplitter yields a bipartite entangled state \cite{Ah}.  
This result is in agreement with entropic measures of 
bipartite pure-state entanglement \cite{Matt}. If one single-mode 
squeezed state were distributed among $N$ parties using linear optics
would we obtain a truly $N$-partite entangled state? We will show that
we can answer this question using the fidelity criterion for teleporting 
unknown coherent states. In particular, we will see that {\it one\/} 
single-mode squeezed state is sufficient to allow quantum teleportation
between any two of the $N$ parties with the help of all other parties.
The assistance by the other $N-2$ parties only relies on local
measurements and classical communication. Due to these $N-2$ measurements,
bipartite entangled states are `distilled' from the initial
$N$-partite entangled state.

The `position' and `momentum' of a 1-D wavepacket 
(units-free with $\hbar=\case{1}{2}$ as in Ref.~\onlinecite{Sam98c})
are the electric quadrature amplitudes
representing the quantum state of a single polarization of a single 
transverse mode of electromagnetic radiation. We define the action 
of an ideal (phase-free) beamsplitter operation on a 
pair of modes $i$ and $j$ via
\begin{equation}\label{BS}
\hat{B}_{ij}(\theta):\left\{
\begin{array}{l}
\hat a_i \rightarrow \hat a_i \cos\theta + \hat a_j\sin\theta \\
\hat a_j \rightarrow \hat a_i \sin\theta - \hat a_j\cos\theta \\
\end{array} \right. \;.
\end{equation}
It yields an entangled state when applied to the zero-momentum eigenstate
$|p=0\rangle\propto\int dx\,|x\rangle$ of mode 1 and the 
zero-position eigenstate $|x=0\rangle$ of mode 2:
\begin{equation}\label{2modeSch}
\hat{B}_{12}(\pi/4)\int dx\,|x,0\rangle\propto \int dx\,|x,x\rangle\;.
\end{equation}
The outgoing EPR state \cite{EPR}, a two-mode momentum 
eigenstate with total momentum $p_1+p_2=0$ and relative position 
$x_1-x_2=0$, contains exactly the correlations we need for reliable 
teleportation. It corresponds to a two-mode squeezed state \cite{Walls},
obtained by superimposing a momentum-squeezed and a position-squeezed 
state, in the limit of infinite squeezing.
We use the Heisenberg representation to describe an approximate version
of this state for finite squeezing. 
The beamsplitter operation applied to a momentum-squeezed and 
a position-squeezed vacuum mode yields for the Heisenberg operators
\cite{Walls}
\begin{eqnarray}\label{2modeHeis}
\hat{x}_1&=&(e^{+r_1} \hat{x}^{(0)}_1+ e^{-r_2} \hat{x}^{(0)}_2)/\sqrt{2} \;,
\nonumber\\
\hat{p}_1&=&(e^{-r_1} \hat{p}^{(0)}_1+ e^{+r_2} \hat{p}^{(0)}_2)/\sqrt{2} \;,
\nonumber\\
\hat{x}_2&=&(e^{+r_1} \hat{x}^{(0)}_1- e^{-r_2} \hat{x}^{(0)}_2)/\sqrt{2} \;,
\nonumber\\
\hat{p}_2&=&(e^{-r_1} \hat{p}^{(0)}_1- e^{+r_2} \hat{p}^{(0)}_2)/\sqrt{2} \;.
\end{eqnarray}
A superscript `$(0)$' denotes initial vacuum modes and $r_1,r_2$ are the 
squeezing parameters. 
For quantum teleportation \cite{Sam98a},
mode 1 is sent to ``Alice'' (the sender) and mode 2 is sent to ``Bob'' (the 
receiver). Alice's mode is superimposed at a 50/50 beamsplitter with the 
unknown input mode ``in'' to be teleported, yielding for the relevant 
quadratures:
$\hat{x}_{\rm u}=(\hat{x}_{\rm in}- \hat{x}_1)/\sqrt{2}$,
$\hat{p}_{\rm v}=(\hat{p}_{\rm in}+ \hat{p}_1)/\sqrt{2}$.
Alice measures certain classical values 
$x_{\rm u}$ and $p_{\rm v}$ for $\hat{x}_{\rm u}$ and $\hat{p}_{\rm v}$.
The operators $\hat{x}_{\rm u}$ and $\hat{p}_{\rm v}$ collapse in
Bob's mode 2 written as
$\hat{x}_2=\hat{x}_{\rm in}-(\hat{x}_1-\hat{x}_2)
-\sqrt{2}\hat{x}_{\rm u}$,
$\hat{p}_2=\hat{p}_{\rm in}+(\hat{p}_1+\hat{p}_2)
-\sqrt{2}\hat{p}_{\rm v}$.
After receiving Alice's classical results $x_{\rm u}$ and $p_{\rm v}$, 
Bob displaces his mode correspondingly, 
$\hat{x}_2\longrightarrow\hat{x}_{\rm tel}=\hat{x}_2+g\,\sqrt{2}
x_{\rm u}$,
$\hat{p}_2\longrightarrow\hat{p}_{\rm tel}=\hat{p}_2+g\,\sqrt{2}
p_{\rm v}$.
The parameter $g$ describes a normalized gain. For $g=1$, the teleported  
mode becomes
$\hat{x}_{\rm tel}=\hat{x}_{\rm in}-\sqrt{2}e^{-r_2} \hat{x}^{(0)}_2$,
$\hat{p}_{\rm tel}=\hat{p}_{\rm in}+\sqrt{2}e^{-r_1} \hat{p}^{(0)}_1$.
Now we assume an arbitrary coherent-state input 
$\alpha_{\rm in}=x_{\rm in}+ip_{\rm in}$ and calculate the 
teleportation fidelity, in this case
defined by $F\equiv\langle\alpha_{\rm in}|\hat{\rho}_{\rm tel}|
\alpha_{\rm in}\rangle$ \cite{Fuchs}.
It describes the overlap between the input and the teleported state
$\hat{\rho}_{\rm tel}$. Up to a factor $\pi$, this fidelity is related to 
the $Q$ function of the teleported mode
[$F=\pi Q_{\rm tel}(\alpha_{\rm in})$]:
\begin{equation}\label{fid2}
F=\frac{1}{2\sqrt{\sigma_x\sigma_p}}\exp\left
[-(1-g)^2\left(\frac{x_{\rm in}^2}{2\sigma_x}
+\frac{p_{\rm in}^2}{2\sigma_p}\right)\right]\;, 
\end{equation}
where $\sigma_x$ and $\sigma_p$ are the variances of the
Q function of the teleported mode for the corresponding quadratures.   
An average fidelity $F_{\rm av}>\case{1}{2}$ (averaged upon the complex 
plane) is only achievable using entanglement \cite{Fuchs}.
For $g=1$, in fact, `classical teleportation' with $r_1=r_2=0$ yields
$F_{\rm av}=F=\case{1}{2}$ for a coherent input and 
$\langle\Delta\hat{x}^2\rangle
_{\rm vacuum}=\langle\Delta\hat{p}^2\rangle_{\rm vacuum}=\case{1}{4}$. 
However, for any $r_1>0$ {\it or\/} $r_2>0$ we obtain 
$F>\case{1}{2}$ with $g=1$. 
We must conclude that the Gaussian two-mode 
state obtained by superimposing {\it one\/} single-mode squeezed state with 
vacuum (any $r_1>0,r_2=0$ or vice versa) exhibits bipartite entanglement. 
Of course, reliable teleportation with perfect fidelity 
$F=1$ (for $g=1$) requires $r_1\to\infty$ {\it and\/} 
$r_2\to\infty$ and hence {\it two\/} single-mode 
squeezed states superimposed. We set $r_2=0$ and look for the maximum fidelity
of coherent-state teleportation achievable with one single-mode squeezed 
state $r_1>0$ as entanglement source. For infinite squeezing
$r_1\to\infty$ and $r_2=0$, we find $F=1/\sqrt{2}$ 
($g=1$).

Here, a non-classical teleportation fidelity serves as {\it sufficient}
criterion for the presence of entanglement. Indeed, here, the violation of
$F\leq \case{1}{2}$ in coherent-state teleportation is consistent 
with the violation
of $\langle(\hat{x}_1-\hat{x}_2)^2\rangle+\langle(\hat{p}_1+
\hat{p}_2)^2\rangle \geq 1$ which has been recently identified as
a sufficient inseparability criterion for bipartite continuous-variable
systems \cite{Duan}. A simpler but less compelling method than doing
quantum teleportation for the experimental application of this sufficient
criterion would be the detection of the variances
$\langle(\hat{x}_1-\hat{x}_2)^2\rangle$ and $\langle(\hat{p}_1+
\hat{p}_2)^2\rangle$ after combining the two modes at a beamsplitter
\cite{Tan}.

We now ask if it is also possible to connect more than two locations
via EPR channels that can be used for quantum teleportation.
At first we consider three locations represented by ``Alice'',
``Bob'' and ``Claire''. Applying the beamsplitter operations 
(``tritter'' \cite{Sam98c})
\begin{equation}\label{tritt}
\hat{T}_{123}\equiv\hat{B}_{23}(\pi/4)\hat{B}_{12}\left(\cos^{-1}1/
\sqrt{3}\right) \;,
\end{equation}
to a zero-momentum eigenstate in mode 1 and a pair of zero-position 
eigenstates in modes 2 and 3 yields $\int dx\,|x,x,x\rangle$.
This GHZ-like state \cite{GHZ} is an eigenstate of total momentum zero
with relative positions $x_i-x_j=0$ $(i,j=1,2,3)$.
It obviously exhibits tripartite entanglement.
In order to consider finite squeezing, we again employ the Heisenberg 
representation. The tritter applied to a momentum-squeezed 
and two position-squeezed vacuum modes yields for the Heisenberg operators
\begin{eqnarray}\label{3modeHeis}
\hat{x}_1&=&\frac{1}{\sqrt{3}}e^{+r_1} \hat{x}^{(0)}_1+\sqrt{\frac{2}{3}}
e^{-r_2} \hat{x}^{(0)}_2\;,\nonumber\\
\hat{p}_1&=&\frac{1}{\sqrt{3}}e^{-r_1} \hat{p}^{(0)}_1+\sqrt{\frac{2}{3}}
e^{+r_2} \hat{p}^{(0)}_2\;,\nonumber\\
\hat{x}_2&=&\frac{1}{\sqrt{3}}e^{+r_1} \hat{x}^{(0)}_1-\frac{1}{\sqrt{6}}
e^{-r_2} \hat{x}^{(0)}_2+\frac{1}{\sqrt{2}} e^{-r_3} \hat{x}^{(0)}_3\;,
\nonumber\\
\hat{p}_2&=&\frac{1}{\sqrt{3}}e^{-r_1} \hat{p}^{(0)}_1-\frac{1}{\sqrt{6}}
e^{+r_2} \hat{p}^{(0)}_2+\frac{1}{\sqrt{2}} e^{+r_3} \hat{p}^{(0)}_3\;,
\nonumber\\
\hat{x}_3&=&\frac{1}{\sqrt{3}}e^{+r_1} \hat{x}^{(0)}_1-\frac{1}{\sqrt{6}}
e^{-r_2} \hat{x}^{(0)}_2-\frac{1}{\sqrt{2}} e^{-r_3} \hat{x}^{(0)}_3\;,
\nonumber\\
\hat{p}_3&=&\frac{1}{\sqrt{3}}e^{-r_1} \hat{p}^{(0)}_1-\frac{1}{\sqrt{6}}
e^{+r_2} \hat{p}^{(0)}_2-\frac{1}{\sqrt{2}} e^{+r_3} \hat{p}^{(0)}_3\;,
\end{eqnarray}
with the three squeezing parameters $r_1$, $r_2$ and $r_3$.
The teleportation protocol involving three participants 
Alice, Bob and Claire works as follows. Let us send the three modes of 
Eqs.~(\ref{3modeHeis}) to Alice, Bob and Claire respectively. 
Again, Alice wants to teleport an unknown quantum
state and couples her mode 1 with the unknown input mode ``in'':
$\hat{x}_{\rm u}=(\hat{x}_{\rm in}- \hat{x}_1)/\sqrt{2}$,
$\hat{p}_{\rm v}=(\hat{p}_{\rm in}+ \hat{p}_1)/\sqrt{2}$.
Let us write Bob's mode 2 and Claire's mode 3 as
\begin{eqnarray}\label{mode2and3}
\hat{x}_2&=&\hat{x}_{\rm in}-(\hat{x}_1-\hat{x}_2)
-\sqrt{2}\hat{x}_{\rm u}\;,\nonumber\\
\hat{p}_2&=&\hat{p}_{\rm in}+(\hat{p}_1+\hat{p}_2+g^{(3)}\hat{p}_3)
-\sqrt{2}\hat{p}_{\rm v}-g^{(3)}\hat{p}_3\;,\nonumber\\
\hat{x}_3&=&\hat{x}_{\rm in}-(\hat{x}_1-\hat{x}_3)
-\sqrt{2}\hat{x}_{\rm u}\;,\nonumber\\
\hat{p}_3&=&\hat{p}_{\rm in}+(\hat{p}_1+g^{(3)}\hat{p}_2+\hat{p}_3)
-\sqrt{2}\hat{p}_{\rm v}-g^{(3)}\hat{p}_2\;,
\end{eqnarray}
where $g^{(3)}$ is another gain determined later.
Again, Alice measures certain classical values 
$x_{\rm u}$ and $p_{\rm v}$ for $\hat{x}_{\rm u}$ and $\hat{p}_{\rm v}$.
The operators $\hat{x}_{\rm u}$ and $\hat{p}_{\rm v}$ in 
Eqs.~(\ref{mode2and3}) collapse. However, this time
Alice sends her classical results $x_{\rm u}$ and $p_{\rm v}$ either to
Bob or Claire via classical channels. Now either Bob or Claire is able to
reconstitute the input state provided that additional classical information
is received: Bob needs the result of a momentum-detection by Claire 
reducing $\hat{p}_3$ to $p_3$ and Claire needs the result of a 
momentum-detection by Bob reducing $\hat{p}_2$ to $p_2$.
Assuming that Claire detects her mode 3 and sends the result to 
Bob, a displacement of Bob's mode 2, 
$\hat{x}_2\longrightarrow\hat{x}_{\rm tel}=\hat{x}_2+g\,\sqrt{2}
x_{\rm u}$,
$\hat{p}_2\longrightarrow\hat{p}_{\rm tel}=\hat{p}_2+g\,\sqrt{2}
p_{\rm v}+g^{(3)}\,p_3$,
accomplishes the teleportation. For $g=1$, the teleported mode becomes
\begin{eqnarray}\label{gain1}
\hat{x}_{\rm tel}&=& \hat{x}_{\rm in} -(\sqrt{3}e^{-r_2} 
\hat{x}^{(0)}_2-e^{-r_3} \hat{x}^{(0)}_3)/\sqrt{2}\;,\\
\hat{p}_{\rm tel}&=&
\hat{p}_{\rm in}+(2+g^{(3)})e^{-r_1} \hat{p}^{(0)}_1/\sqrt{3} \nonumber\\ 
&+&(1-g^{(3)})e^{+r_2} \hat{p}^{(0)}_2/\sqrt{6} 
+(1-g^{(3)})e^{+r_3} \hat{p}^{(0)}_3/\sqrt{2} \nonumber\;.
\end{eqnarray}
When $r_1=r_2=r_3=r$, the optimum teleportation fidelity is achieved 
with $g^{(3)}=(e^{+4r}-1)/(e^{+4r}+1/2)$
and becomes for a coherent-state input 
with Eqs.~(\ref{gain1}) according to Eq.~(\ref{fid2}) for $g=1$
($F=F_{\rm av}$)
\begin{equation}\label{fid4}
F_{\rm opt}=\left\{\left[1+e^{-2r}\right]
\left[1+3/(2e^{+2r}+e^{-2r})\right]\right\}^{-1/2}
\!\!\!.
\end{equation}
For $r=0$, we obtain $F_{\rm opt}=\case{1}{2}$. Perfect teleportation with 
fidelity $F_{\rm opt}=1$ is achieved for infinite squeezing in 
{\it all three\/}
single-mode squeezed states $r\to\infty$ ($g^{(3)}=1$).
For any $r>0$, we find $F_{\rm opt}>\case{1}{2}$.
However, again,
$F>\case{1}{2}$ can even be satisfied using only {\it one\/}
single-mode squeezed state and two vacua. In this case ($r_2=r_3=0$),
we obtain the optimum fidelity 
$F_{\rm opt}=\left[2+6/(1+2e^{+2r_1})\right]^{-1/2}$
with $g^{(3)}=(e^{+2r_1}-1)/(e^{+2r_1}+1/2)$.
Remarkably, still $F_{\rm opt}>\case{1}{2}$ for any $r_1>0$.
If Alice and Bob arrange to teleport Alice's unknown
coherent state to Claire (and both send the required classical information 
to Claire and Claire performs the corresponding displacements), one can
easily see that with optimum gain the fidelity also exceeds the classical 
limit for any $r_1>0$ when $r_2=r_3=0$. In fact,
Alice, Bob and Claire can determine any one of them as sender and another 
one as receiver.
For $r_1,r_2,r_3\to\infty$ and unit gain, quantum teleportation
is perfect with $F=1$. If $r_1=r_2=r_3=r$, coherent-state teleportation 
with $F>\case{1}{2}$ between any two of Alice, Bob and Claire is achieved 
for any $r>0$, provided optimum gain is used.
Even if the tripartite entanglement is based only on one squeezed
state, the teleportation is better 
than classical with any sender and receiver chosen and any nonzero 
squeezing.
For $r_1\to\infty$ ($r_2=r_3=0$), 
we find the maximum fidelity $F=1/\sqrt{2}$ as in the scheme involving 
only Alice and Bob.

In the following we want to investigate if the previous results can be 
extended to more than three parties.
We apply the beamsplitter operations 
\begin{eqnarray}\label{Nsplitt}
\hat{N}_{1\ldots N}&\equiv&\hat{B}_{N-1\,N}(\pi/4)\hat{B}_{N-2\,N-1}
\left(\cos^{-1}1/\sqrt{3}\right)\nonumber\\
&&\times\cdots\times\hat{B}_{12}\left(\cos^{-1}1/\sqrt{N}\right) \;,
\end{eqnarray}
to a zero-momentum eigenstate in mode 1 and $N-1$ zero-position 
eigenstates in modes $2$ through $N$.
We obtain the entangled $N$-mode state $\int dx\,|x,x,\ldots ,x\rangle$.
This state is an eigenstate with total momentum zero and relative 
positions $x_i-x_j=0$ $(i,j=1,2,\ldots ,N)$.

For finite squeezing, again we refer to the Heisenberg operators. The 
above ``$N$-splitter'' applied to one momentum-squeezed ($r_1=r$)
and $N-1$ position-squeezed vacuum modes ($r_2=r_3=\cdots =r_N=r$) 
yields momentum quadrature operators for any $N$ correlated as
\begin{eqnarray}\label{corr3}
&&\left\langle\left(\hat{p}_k+\hat{p}_l+g^{(N)}\sum^N_{j\neq k,l}
\hat{p}_j\right)^2\right\rangle=\\
&&\frac{[2+(N-2)g^{(N)}]^2}{4N}e^{-2r}
+\frac{(g^{(N)}-1)^2(N-2)}{2N}e^{+2r}\nonumber\;,
\end{eqnarray}
which becomes zero and means perfect correlations for $r\to\infty$ 
and $g^{(N)}=1$ ($k\neq l$).  
The correlations of the outgoing position quadrature operators satisfy
$\langle(\hat{x}_k-\hat{x}_l)^2\rangle=e^{-2r}/2$.
The symmetric $N$-mode Wigner function of these states is a 
generalization of the finite-squeezing EPR-state Wigner 
function \cite{PvL}. 
With only one momentum-squeezed (squeezing $r_1$) and
$N-1$ vacuum modes ($r_2=r_3=\cdots =r_N=0$), the variance
in Eq.~(\ref{corr3}) becomes
$[2+(N-2)g^{(N)}]^2e^{-2r_1}/(4N)+(g^{(N)}-1)^2
(N-2)/(2N)$ and $\langle(\hat{x}_k-\hat{x}_l)^2\rangle=1/2$. 

Let us now assume that the $N$ outgoing modes are sent to $N$ different
locations. We arbitrarily choose two locations of them as sending (mode $k$)
and receiving station (mode $l$) for teleportation. The teleportation 
protocol is almost the same as in the $N$=3-case. 
However, now the receiver needs the classical information
of the sender's detection of the quadratures
$\hat{x}_{\rm u}=(\hat{x}_{\rm in}- \hat{x}_k)/\sqrt{2}$,
$\hat{p}_{\rm v}=(\hat{p}_{\rm in}+ \hat{p}_k)/\sqrt{2}$,
and in addition the classical results 
of $N-2$ momentum-detections at the $N-2$ remaining stations.
This can be seen by writing mode $l$ as
\begin{eqnarray}\label{model}
\hat{x}_l&=&\hat{x}_{\rm in}-(\hat{x}_k-\hat{x}_l)
-\sqrt{2}\hat{x}_{\rm u}\;,\\
\hat{p}_l&=&\hat{p}_{\rm in}+
\hat{p}_k+\hat{p}_l+g^{(N)}\sum^N_{j\neq k,l}\hat{p}_j
-\sqrt{2}\hat{p}_{\rm v}-g^{(N)}\sum^N_{j\neq k,l}\hat{p}_j\nonumber\;.
\end{eqnarray}
Finally, the receiver displaces his mode by the sum of all classical results  
received,
$\hat{x}_l\longrightarrow\hat{x}_{\rm tel}=\hat{x}_l+g\,\sqrt{2}
x_{\rm u}$,
$\hat{p}_l\longrightarrow\hat{p}_{\rm tel}=\hat{p}_l+g\,\sqrt{2}
p_{\rm v}+g^{(N)}\sum^N_{j\neq k,l}p_j$.
For $g=1$, the teleported mode becomes
$\hat{x}_{\rm tel}=\hat{x}_{\rm in}-(\hat{x}_k-\hat{x}_l)$,
$\hat{p}_{\rm tel}=\hat{p}_{\rm in}+
\hat{p}_k+\hat{p}_l+g^{(N)}\sum^N_{j\neq k,l}\hat{p}_j$.

\begin{figure}[htb]
\begin{center}
\begin{psfrags}
     \psfrag{F}[cb]{\Large $F~~~$}
     \psfrag{squeezing}[c]{\Large ~~~~~~~~squeezing [dB]}
     \psfrag{N=2}[c]{$~~~~~~~~2$}
     \psfrag{N=3}[c]{$~~~3$}
     \psfrag{N=4}[c]{$~~~~~4$}
     \psfrag{N=8}[c]{$8$}
     \psfrag{N=20}[c]{$~~20$}
     \psfrag{N=50}[c]{$50~$}
     \psfrag{classical}{\Large \bf classical}
     \psfrag{quantum}{\Large \bf quantum}
     \psfrag{N squeezed states}{}%\Large \bf N squeezed states}
\epsfxsize=3.2in
     \epsfbox[0 60 400 320]{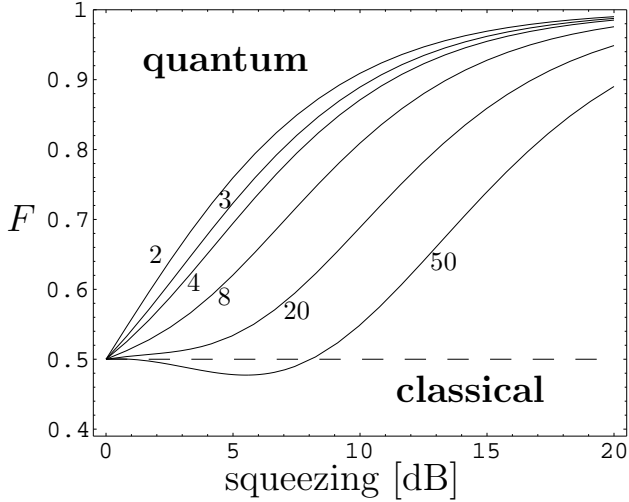}
\end{psfrags}
\end{center}
\caption{$N$ squeezed states:
         Optimized fidelity for the teleportation of an arbitrary 
         coherent state from any sender to any receiver chosen 
         from $N$ ($=2$, $3$, $4$, $8$, $20$ and $50$) parties.
         A fidelity $F>0.5$ (``quantum'') requires $N$-partite
         entanglement, here produced with $N$ 
         equally squeezed single-mode states. For $N\geq 30$, 
         the fidelity of our protocol becomes classical
         for some squeezing, but always exceeds $0.5$ for sufficiently 
         {\it small} squeezing and approaches 1 for
         infinite squeezing.} 
\label{fig1}
\end{figure}

Now we can optimize the teleportation fidelity 
using Eq.~(\ref{corr3}) and find the optimum gain
$g^{(N)}=[e^{+4r}-1]/[e^{+4r}+(N-2)/2]$, assuming $r_1=r_2=\cdots =r_N=r$.
For a coherent-state input, 
we obtain the optimum fidelity according to Eq.~(\ref{fid2}) with $g=1$
($F=F_{\rm av}$)
\begin{eqnarray}\label{fidelity}
F_{\rm opt}&=&\left\{1+e^{-2r}\right\}^{-1/2}\nonumber\\
&\times&\left\{1+N/[2e^{+2r}+(N-2)e^{-2r}]\right\}^{-1/2}
\!\!\!.
\end{eqnarray}
For $r=0$, we obtain $F_{\rm opt}=\case{1}{2}$. Perfect teleportation with 
$F_{\rm opt}=1$ in any of the $N(N-1)/2$ channels 
(but, of course, not simultaneously by no-cloning \cite{Woott}) 
is achieved with infinite squeezing in {\it all\/}
single-mode squeezed states $r\to\infty$ ($g^{(N)}=1$)
for any sending and receiving station chosen from the $N$ 
locations.
For any $r>0$, we find $F_{\rm opt}>\case{1}{2}$,
{\it provided $N\leq 29$}.
Interestingly, if $N\geq 27$, $F_{\rm opt}$ reaches a maximum and then
drops to a minimum before approaching 1 when the squeezing is increased.
For $N\geq 30$, the minimum is below $\case{1}{2}$, but the maximum
(at sufficiently small squeezing) still exceeds $\case{1}{2}$.
Figure~\ref{fig1} shows the fidelity of Eq.(\ref{fidelity}) 
for squeezing in dB.

% FIG 1

\begin{figure}[htb]
\begin{center}
\begin{psfrags}
     \psfrag{F}[cb]{\Large $F~~~$}
     \psfrag{squeezing}[c]{\Large ~~~~~~~~squeezing [dB]}
     \psfrag{N=2}[c]{$~~~~~~~~2$}
     \psfrag{N=3}[c]{$~3$}
     \psfrag{N=4}[cb]{$~~~~4$}
     \psfrag{N=8}[c]{$8$}
     \psfrag{N=20}[c]{$~~20$}
     \psfrag{N=50}[c]{$50~$}
     \psfrag{classical}{\Large \bf classical}
     \psfrag{quantum}{\Large \bf quantum}
     \psfrag{one squeezed state}{}%\Large \bf one squeezed state}
\epsfxsize=3.2in
     \epsfbox[0 60 400 320]{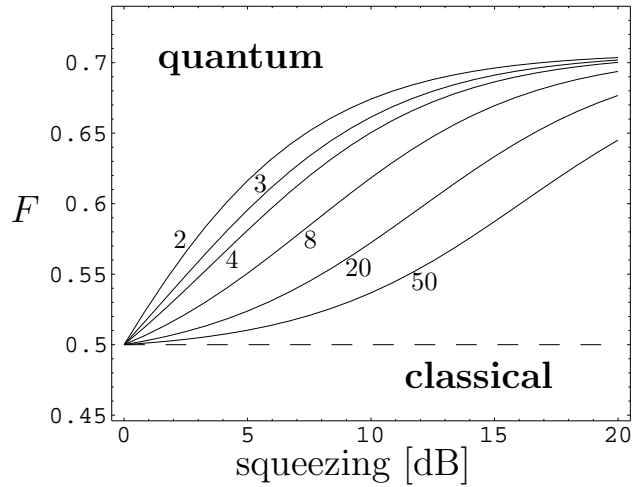}
\end{psfrags}
\end{center}
\caption{One squeezed state: Optimized fidelity for the 
         teleportation of an arbitrary coherent state from
         any sender to any receiver chosen from $N$
         ($=2$, $3$, $4$, $8$, $20$ and $50$) parties.
         A fidelity $F>0.5$ (``quantum'') requires $N$-partite
         entanglement, here produced with 
         {\it one\/} single-mode squeezed state.}
\label{fig2}
\end{figure}

What about using only {\it one\/} single-mode squeezed state and
$N-1$ vacua in this $N$-mode scheme?
Even in this case ($r_2=r_3=\cdots =r_N=0$),
quantum teleportation is possible in any of the
$N(N-1)/2$ channels for any $N$. We obtain the optimum fidelity
for coherent-state teleportation 
$F_{\rm opt}=\left[2+2N/(N-2+2e^{+2r_1})\right]^{-1/2}$
with $g^{(N)}=[e^{+2r_1}-1]/[e^{+2r_1}+(N-2)/2]$,
shown in Fig.~\ref{fig2} for squeezing in dB.
Remarkably, $F_{\rm opt}>\case{1}{2}$ for any $r_1>0$ with 
{\it arbitrary $N$}.
In the limit $r_1\to\infty$, we still attain the maximum fidelity
$F=1/\sqrt{2}$ for any $N$. 

By first considering only the momentum-detections at the $N-2$ 
stations without the teleportation from $k$ to $l$, we can give our protocol
also the quality of a `distillation' of bipartite entanglement from 
$N$-partite entanglement. The bipartite entanglement
can experimentally be verified by applying sufficient inseparability
criteria through detections of the combined modes \cite{Tan}
or through quantum teleportation as shown.
These verifications require classical communication and local 
displacements based on the $N-2$ measurement results.
However, both in the scheme with $N$ 
squeezed states and with one squeezed state, mode $k$ and $l$
are projected on bipartite entangled states for any
nonzero squeezing and arbitrary $N$ just due to the collapses
of the $N-2$ momenta \cite{PvL}.
This indicates that in particular our scheme with $N$  
squeezed states, yielding classical fidelities for $N\geq 30$ and 
some squeezing, might not be optimal. Yet asymmetric displacements
by the $N-2$ classical results do not provide better fidelities. 

% FIG 2

In summary, we have considered multipartite
entanglement based on quantum variables with a continuous spectrum
and a quantum teleportation network using 
this multipartite entanglement. It can be comparatively easily
generated using squeezed light and linear optics. 

This work was funded by a DAAD Doktorandenstipendium (HSP III) 
and by the EPSRC Grant No.\ GR/L91344.
SLB thanks Eugene Polzik for suggesting 
we consider teleportation with non-symmetric squeezing.

\end{document}